\documentclass[12pt]{article}
\begin{document}

\author{C. Bizdadea\thanks{%
e-mail address: bizdadea@central.ucv.ro}, E.\ M. Cioroianu\thanks{%
e-mail address: manache@central.ucv.ro}, S. O. Saliu\thanks{%
e-mail address: osaliu@central.ucv.ro} \\
Faculty of Physics, University of Craiova\\
13 A. I. Cuza Str., Craiova RO-1100, Romania}
\title{Irreducible Freedman-Townsend vertex and Hamiltonian BRST cohomology}
\maketitle

\begin{abstract}
The irreducible Freedman-Townsend vertex is derived by means of the
Hamiltonian deformation procedure based on local BRST cohomology.

PACS number: 11.10.Ef
\end{abstract}

\section{Introduction}

The key point in the development of the BRST formalism \cite{1}--\cite{13}
is the recursive pattern of homological perturbation theory \cite{14}--\cite
{20}. The Hamiltonian BRST formulation was proved to be a natural background
for analysing various topics, such as the construction of such a symmetry in
quantum mechanics \cite{8} (Chapter 14), or an appropriate correlation
between the BRST symmetry itself and canonical quantization methods \cite{21}%
. Its cohomological understanding helped at clarifying some new aspects,
like the Hamiltonian analysis of anomalies \cite{22}, the precise relation
between local Lagrangian and Hamiltonian BRST cohomologies \cite{23}, the
construction of irreducible BRST symmetries for reducible theories \cite{24}%
, or the study of consistent Hamiltonian interactions that can be added
among fields with gauge freedom \cite{25}--\cite{28}.

In this paper we investigate the consistent interactions that can be
introduced among a system of two-form gauge fields in four dimensions by
means of an irreducible Hamiltonian BRST deformation procedure, in spite of
the reducibility of the initial model. Our strategy goes as follows. First,
we start with a system of abelian two-forms in four dimensions, described by
a reducible BRST differential, $s_{R}$, to whom we associate an irreducible
BRST differential $s_{I}$, with the properties 
\begin{equation}
s_{R}^{2}=0=s_{I}^{2},\;H^{0}\left( s_{R}\right) \simeq H^{0}\left(
s_{I}\right) ,  \label{basic}
\end{equation}
where $H^{k}\left( s\right) $ denotes the $k$th order cohomological space of
the differential $s$. The relations (\ref{basic}) allow us to replace the
reducible BRST differential with the irreducible one from the point of view
of the basic equations of the BRST formalism. Second, we apply the
Hamiltonian deformation technique to the irreducible version of the model
under study. It has been shown in \cite{25}--\cite{28} that the Hamiltonian
problem of constructing consistent interactions can be reformulated as a
deformation problem of the BRST charge (canonical generator of the BRST
symmetry) and BRST-invariant Hamiltonian of a given free theory. In turn,
the Hamiltonian deformation scheme reduces to two towers of equations
involving the free BRST differential. The free irreducible BRST differential
splits in our case as $s_{I}=\delta _{I}+\gamma _{I}$, where $\delta _{I}$
stands for the Koszul-Tate differential, and $\gamma _{I}$ represents the
exterior derivative along the gauge orbits. In order to generate the
first-order deformation of the non-integrated BRST charge density, we
perform its expansion according to an auxiliary degree, called antighost
number, and assume that we can take the last representative of this
expansion to be annihilated by $\gamma _{I}$. In consequence, we have to
know $H\left( \gamma _{I}\right) $. In the meantime, the computation of the
before last term of this expansion requires the knowledge of $H\left( \delta
_{I}|\tilde{d}\right) $, where $\tilde{d}$ is the spatial part of the
exterior space-time derivative, $\tilde{d}=dx^{i}\partial _{i}$. After the
computation of these cohomologies, we appropriately solve the deformation
equations, finally obtaining the BRST charge and BRST-invariant Hamiltonian
of the interacting theory. Third, from these quantities we identify the
structure of the deformed gauge theory, hence the first-class constraints,
the first-class Hamiltonian, and the accompanying gauge algebra. The
resulting model is precisely the irreducible version of the non-abelian
Freedman-Townsend model \cite{29}. It is well-known that the
Freedman-Townsend model plays a special role due on the one hand to its link
with Witten's string theory \cite{30}, and, on the other hand, to its
equivalence to the non-linear $\sigma $-model \cite{29}.

Our paper is structured in five sections. Section 2 is devoted to the
construction of an irreducible Hamiltonian BRST differential for a
four-dimensional system of abelian two-forms. In Section 3 we deal with the
deformation of this irreducible BRST symmetry on account of cohomological
techniques. Section 4 realizes the identification of the resulting deformed
model, while Section 5 ends the paper with some conclusions.

\section{Irreducible BRST differential of the free model}

In this section we construct an irreducible BRST differential for a set of
abelian two-forms in four dimensions.

\subsection{The undeformed model}

We begin with the first-order Lagrangian action that describes a system of
free two-forms 
\begin{equation}
S_{0}^{L}\left[ A_{\mu }^{a},B_{a}^{\mu \nu }\right] =\frac{1}{2}\int
d^{4}x\left( -B_{a}^{\mu \nu }F_{\mu \nu }^{a}+A_{\mu }^{a}A_{a}^{\mu
}\right) ,  \label{1}
\end{equation}
where $B_{a}^{\mu \nu }$ stands for a set of antisymmetric tensor fields,
and the field strength of $A_{\mu }^{a}$ reads as $F_{\mu \nu }^{a}=\partial
_{\mu }A_{\nu }^{a}-\partial _{\nu }A_{\mu }^{a}$. Eliminating the
second-class constraints with the help of the Dirac bracket built with
respect to themselves (the independent `co-ordinates' of the reduced
phase-space are $\bar{z}^{A}=\left( A_{i}^{a},B_{a}^{0i},B_{a}^{ij},\pi
_{ij}^{a}\right) $), we obtain the first-class constraints 
\begin{equation}
\gamma _{1i}^{a}\equiv \epsilon _{0ijk}\pi ^{ajk}\approx
0,\;G_{2i}^{a}\equiv \frac{1}{2}\epsilon _{0ijk}F^{ajk}\approx 0,  \label{9}
\end{equation}
and the first-class Hamiltonian 
\begin{equation}
H_{0}^{\prime }=\frac{1}{2}\int d^{3}x\left(
B_{a}^{ij}F_{ij}^{a}-A_{i}^{a}A_{a}^{i}+\left( \partial
^{i}B_{0i}^{a}\right) \left( \partial _{j}B_{a}^{0j}\right) \right) \equiv
\int d^{3}x\,h_{0}^{\prime },  \label{10}
\end{equation}
where the non-vanishing Dirac brackets among the independent components are
expressed by 
\begin{equation}
\left[ B_{a}^{0i}\left( x\right) ,A_{j}^{b}\left( y\right) \right]
_{x^{0}=y^{0}}^{*}=\delta _{a}^{\;\;b}\delta _{\;\;j}^{i}\delta ^{3}\left( 
\mathbf{x}-\mathbf{y}\right) ,  \label{7}
\end{equation}
\begin{equation}
\left[ B_{a}^{ij}\left( x\right) ,\pi _{kl}^{b}\left( y\right) \right]
_{x^{0}=y^{0}}^{*}=\frac{1}{2}\delta _{a}^{\;\;b}\delta _{\;\;k}^{\left[
i\right. }\delta _{\;\;l}^{\left. j\right] }\delta ^{3}\left( \mathbf{x}-%
\mathbf{y}\right) .  \label{8}
\end{equation}
We remark that the functions $G_{2i}^{a}$ from (\ref{9}) are not independent
(first-stage reducible) 
\begin{equation}
\partial ^{i}G_{2i}^{a}=0,  \label{11}
\end{equation}
while the entire first-class constraint set (\ref{9}) remains abelian in
terms of the Dirac bracket.

\subsection{Irreducible Hamiltonian BRST symmetry}

Here, we derive an irreducible Hamiltonian BRST symmetry associated with the
model under study by following the general line exposed in \cite{24}. Thus,
for a system subject to the first-class constraints 
\begin{equation}
G_{a_{0}}\left( \bar{z}^{A}\right) \approx 0,\;a_{0}=1,\ldots ,M_{0},
\label{11x}
\end{equation}
that are first-stage reducible 
\begin{equation}
Z_{\;\;a_{1}}^{a_{0}}G_{a_{0}}=0,\;a_{1}=1,\ldots ,M_{1},  \label{11y}
\end{equation}
we can construct a corresponding theory, based on the irreducible
first-class constraints 
\begin{equation}
\gamma _{a_{0}}\equiv G_{a_{0}}+A_{a_{0}}^{\;\;a_{1}}\pi _{a_{1}}\approx 0,
\label{11z}
\end{equation}
where $\left( y^{a_{1}},\pi _{a_{1}}\right) $ denote some new canonical
pairs that extend the original phase-space, and $A_{a_{0}}^{\;\;a_{1}}$ are
some functions (that may involve at most the original variables $\bar{z}^{A}$%
), taken to satisfy the condition 
\begin{equation}
rank\left( Z_{\;\;a_{1}}^{a_{0}}A_{a_{0}}^{\;\;b_{1}}\right) =M_{1}.
\label{11xy}
\end{equation}
In order to infer an irreducible Lagrangian formulation that is manifestly
covariant, it is necessary to add some supplementary canonical pairs $\left(
y^{\left( 1\right) a_{1}},\pi _{a_{1}}^{\left( 1\right) }\right) $ and $%
\left( y^{\left( 2\right) a_{1}},\pi _{a_{1}}^{\left( 2\right) }\right) $,
subject to the constraints 
\begin{equation}
\gamma _{1a_{1}}\equiv \pi _{a_{1}}-\pi _{a_{1}}^{\left( 1\right) }\approx
0,\;\gamma _{2a_{1}}\equiv -\pi _{a_{1}}^{\left( 2\right) }\approx 0,
\label{11xz}
\end{equation}
which together with (\ref{11z}) form an irreducible first-class set. Under
these circumstances, it is shown \cite{24} that we can construct the
Hamiltonian BRST symmetry $s_{I}$ of the irreducible theory based on the
first-class constraints (\ref{11z}) and (\ref{11xz}), such that $%
s_{I}^{2}=0=s_{R}^{2}$ and $H^{0}\left( s_{I}\right) \simeq H^{0}\left(
s_{R}\right) $, where $s_{R}$ stands for the Hamiltonian BRST symmetry
within the initial reducible formulation. The last formulas indicate that
the BRST symmetries $s_{I}$ and $s_{R}$ are equivalent from the point of
view of the fundamental equations of the Hamiltonian BRST formalism, namely,
the nilpotency of the BRST operator and the isomorphism between the
zeroth-order cohomological space of the BRST differential and the algebra of
physical observables. As a consequence, it is permissible to replace the
reducible BRST symmetry with the irreducible one. In this light, we further
analyse the construction of the irreducible Hamiltonian BRST symmetry for
abelian two-forms.

Initially, we determine the concrete form of the irreducible first-class
constraints. In this respect, we observe that in the case of the model under
study we have that $Z_{\;\;a_{1}}^{a_{0}}\rightarrow \delta
_{\;\;b}^{a}\partial ^{i}$. If we take $A_{a_{0}}^{\;\;a_{1}}\rightarrow
-\delta _{\;\;c}^{b}\partial _{i}$, then the requirement (\ref{11xy}) is
indeed fulfilled. Accordingly, the irreducible first-class constraints (\ref
{11z}) and (\ref{11xz}) take the concrete form 
\begin{equation}
\gamma _{1i}^{a}\equiv \epsilon _{0ijk}\pi ^{ajk}\approx 0,\;\gamma
_{2i}^{a}\equiv \frac{1}{2}\epsilon _{0ijk}F^{ajk}-\partial _{i}\pi
^{a}\approx 0,  \label{11u}
\end{equation}
\begin{equation}
\gamma _{1}^{a}\equiv \pi ^{a}-\pi ^{(1)a}\approx 0,\;\gamma _{2}^{a}\equiv
-\pi ^{(2)a}\approx 0,  \label{11v}
\end{equation}
where $\left( \varphi _{a},\pi ^{a}\right) $, $\left( \varphi _{a}^{\left(
1\right) },\pi ^{(1)a}\right) $ and $\left( \varphi _{a}^{\left( 2\right)
},\pi ^{(2)a}\right) $ represent the supplementary canonical pairs mentioned
in the above. Thus, the irreducible constraint set preserves the abelian
behaviour (in the Dirac bracket) of the former reducible one. We will work
with a first-class Hamiltonian of the type 
\begin{eqnarray}
& &H_{0}=\int d^{3}x\left( \frac{1}{2}B_{a}^{ij}\left( F_{ij}^{a}+\epsilon
_{0ijk}\partial ^{k}\pi ^{a}\right) -\frac{1}{2}A_{i}^{a}A_{a}^{i}+\right. 
\nonumber \\
& &\left. \varphi _{a}\pi ^{(2)a}+\frac{1}{2}\left( \partial
^{i}B_{0i}^{a}\right) \partial _{j}B_{a}^{0j}+\left( \partial _{i}\varphi
_{a}^{(2)}\right) \partial ^{i}\pi ^{a}\right) ,  \label{12x}
\end{eqnarray}
with the help of which we infer the gauge algebra relations 
\begin{equation}
\left[ H_{0},\gamma _{1i}^{a}\right] ^{*}=\gamma _{2i}^{a},\;\left[
H_{0},\gamma _{2i}^{a}\right] ^{*}=\partial _{i}\gamma _{2}^{a},  \label{13x}
\end{equation}
\begin{equation}
\left[ H_{0},\gamma _{1}^{a}\right] ^{*}=-\gamma _{2}^{a},\;\left[
H_{0},\gamma _{2}^{a}\right] ^{*}=-\partial ^{i}\gamma _{2i}^{a}.
\label{14x}
\end{equation}
We have all the elements necessary at the construction of the main
ingredients of the Hamiltonian BRST formulation of the new model. The BRST
charge and BRST-invariant Hamiltonian of this irreducible free theory are
expressed by 
\begin{equation}
\Omega _{0}=\int d^{3}x\sum\limits_{\Delta =1}^{2}\left( \eta _{\Delta
a}^{i}\gamma _{\Delta i}^{a}+\eta _{\Delta a}\gamma _{\Delta }^{a}\right)
\equiv \int d^{3}x\omega _{0},  \label{15x}
\end{equation}
\begin{equation}
H_{0_{B}}=H_{0}+\int d^{3}x\left( \eta _{1a}^{i}\mathcal{P}_{2i}^{a}-\eta
_{1a}\mathcal{P}_{2}^{a}-\eta _{2a}\partial ^{i}\mathcal{P}_{2i}^{a}+\eta
_{2a}^{i}\partial _{i}\mathcal{P}_{2}^{a}\right) \equiv \int d^{3}xh_{0_{B}},
\label{16x}
\end{equation}
where $\eta ^{\Gamma }\equiv \left( \eta _{1a}^{i},\eta _{2a}^{i},\eta
_{1a},\eta _{2a}\right) $ are fermionic ghost number one ghosts, and $%
\mathcal{P}_{\Gamma }\equiv \left( \mathcal{P}_{1i}^{a},\mathcal{P}_{2i}^{a},%
\mathcal{P}_{1}^{a},\mathcal{P}_{2}^{a}\right) $ denote their corresponding
ghost number minus one antighosts. The ghost number ($\mathrm{gh}$) is
defined like the difference between the pure ghost number ($\mathrm{pgh}$)
and the antighost number ($\mathrm{antigh}$), with 
\begin{equation}
\mathrm{pgh}\left( z^{A}\right) =0,\;\mathrm{pgh}\left( \eta ^{\Gamma
}\right) =1,\;\mathrm{pgh}\left( \mathcal{P}_{\Gamma }\right) =0,
\label{17x}
\end{equation}
\begin{equation}
\mathrm{antigh}\left( z^{A}\right) =0,\;\mathrm{antigh}\left( \eta ^{\Gamma
}\right) =0,\;\mathrm{antigh}\left( \mathcal{P}_{\Gamma }\right) =1,
\label{18x}
\end{equation}
where 
\begin{equation}
z^{A}=\left( A_{i}^{a},B_{a}^{0i},B_{a}^{ij},\pi _{ij}^{a},\varphi _{a},\pi
^{a},\varphi _{a}^{\left( 1\right) },\pi ^{(1)a},\varphi _{a}^{\left(
2\right) },\pi ^{(2)a}\right) .  \label{19x}
\end{equation}

From (\ref{15x}) it is easy to see that the irreducible BRST symmetry $%
s_{I}\bullet =\left[ \bullet ,\Omega _{0}\right] ^{*}$ of the free theory
decomposes like $s_{I}=\delta _{I}+\gamma _{I}$, where $\delta _{I}$ is the
irreducible Koszul-Tate differential, and $\gamma _{I}$ represents the
irreducible exterior derivative along the gauge orbits. These operators act
on the generators of the free irreducible BRST complex through the relations 
\begin{equation}
\delta _{I}z^{A}=0,\;\delta _{I}\eta ^{\Gamma }=0,\;\delta _{I}\mathcal{P}%
_{1i}^{a}=-\epsilon _{0ijk}\pi ^{jka},  \label{20x}
\end{equation}
\begin{equation}
\delta _{I}\mathcal{P}_{2i}^{a}=-\frac{1}{2}\epsilon _{0ijk}F^{jka}+\partial
_{i}\pi ^{a},\;\delta _{I}\mathcal{P}_{1}^{a}=-\pi ^{a}+\pi ^{(1)a},\;\delta
_{I}\mathcal{P}_{2}^{a}=\pi ^{(2)a},  \label{21x}
\end{equation}
\begin{equation}
\gamma _{I}A_{i}^{a}=0,\;\gamma _{I}B_{a}^{0i}=\epsilon ^{0ijk}\partial
_{j}\eta _{2ak},\;\gamma _{I}B_{a}^{ij}=\epsilon ^{0ijk}\eta _{1ak},\;\gamma
_{I}\pi _{ij}^{a}=0,  \label{22x}
\end{equation}
\begin{equation}
\gamma _{I}\varphi _{a}=\eta _{1a}+\partial _{i}\eta _{2a}^{i},\;\gamma
_{I}\varphi _{a}^{\left( 1\right) }=-\eta _{1a},\;\gamma _{I}\varphi
_{a}^{\left( 2\right) }=-\eta _{2a},  \label{23x}
\end{equation}
\begin{equation}
\gamma _{I}\pi ^{a}=\gamma _{I}\pi ^{(1)a}=\gamma _{I}\pi ^{(2)a}=0,\;\gamma
_{I}\eta ^{\Gamma }=0,\;\gamma _{I}\mathcal{P}_{\Gamma }=0.  \label{24x}
\end{equation}
The last definitions will be used in the sequel during the deformation
procedure.

\section{Construction of irreducible deformations}

In this section we determine the consistent interactions that can be added
to action (\ref{1}) in the framework of the irreducible Hamiltonian BRST
background established before. Based on the results from \cite{25}--\cite{28}%
, we can reformulate the general problem of constructing consistent
Hamiltonian interactions as a deformation problem of the BRST charge and
BRST-invariant Hamiltonian associated with a given free theory. If we expand
the BRST charge $\Omega $ of the interacting theory in the powers of the
deformation parameter $g$, $\Omega =\Omega _{0}+g\Omega _{1}+g^{2}\Omega
_{2}+\cdots $, and ask that $\left[ \Omega ,\Omega \right] ^{*}=0$, we find
the equations 
\begin{equation}
\left[ \Omega _{1},\Omega _{0}\right] ^{*}=0,\;\frac{1}{2}\left[ \Omega
_{1},\Omega _{1}\right] ^{*}+\left[ \Omega _{2},\Omega _{0}\right]
^{*}=0,\;\cdots ,  \label{25x}
\end{equation}
where the equation for power zero in $g$ was omitted as it is automatically
obeyed. Let $H_{B}=H_{0_{B}}+gH_{1}+g^{2}H_{2}+\cdots $ be the
BRST-invariant Hamiltonian of the interacting theory. The BRST-invariance of 
$H_{B}$ with respect to $\Omega $ generates the equations 
\begin{equation}
\left[ H_{1},\Omega _{0}\right] ^{*}+\left[ H_{0_{B}},\Omega _{1}\right]
^{*}=0,\;\left[ H_{2},\Omega _{0}\right] ^{*}+\left[ H_{1},\Omega
_{1}\right] ^{*}+\left[ H_{0_{B}},\Omega _{2}\right] ^{*}=0,\;\cdots ,
\label{26x}
\end{equation}
where the zeroth order equation in $g$ is verified by assumption, and was
therefore discarded. Thus, the relations (\ref{25x}-\ref{26x}) completely
describe the BRST approach to the construction of consistent Hamiltonian
interactions, and will accordingly be named the main equations of the
Hamiltonian deformation procedure. The equation that governs the first-order
deformation of the BRST charge (the first relation in (\ref{25x})), demands
that $\Omega _{1}$ should be an $s_{I}$-co-cycle. Trivial co-cycles lead to
trivial deformations (that can be absorbed by a field redefinition), and
will be removed. The second equation in (\ref{25x}) asks that $\left[ \Omega
_{1},\Omega _{1}\right] ^{*}$ should be BRST-exact in order to ensure the
existence of the second-order deformation $\Omega _{2}$.\ In this context,
we are interested in local deformations only, i.e., in the solutions $\Omega
_{k}=\int d^{3}x\omega _{k}$, $H_{k}=\int d^{3}xh_{k}$, with $\omega _{k}$
and $h_{k}$ local functions, so the relevant cohomology space in terms of
the integrands is $H\left( s_{I}|\tilde{d}\right) $.

The first equation in (\ref{25x}) holds if and only if $\omega _{1}$ is a
BRST-co-cycle modulo $\tilde{d}$, i.e., 
\begin{equation}
s_{I}\omega _{1}=\partial _{i}n^{i},  \label{28x}
\end{equation}
for some $n^{i}$. In order to solve (\ref{28x}), we expand $\omega _{1}$
according to the antighost number, $\omega _{1}=\stackrel{\left( 0\right) }{%
\omega }_{1}+\stackrel{\left( 1\right) }{\omega }_{1}+\cdots +\stackrel{%
\left( J\right) }{\omega }_{1}$, where the last term can be assumed to be
annihilated by $\gamma _{I}$, so $\gamma _{I}\stackrel{\left( J\right) }{%
\omega }_{1}=0$. Then, in order to compute the first-order deformation of
the BRST charge, we need to know $H\left( \gamma _{I}\right) $. Analysing
the definitions (\ref{22x}-\ref{24x}), we remark that the fields $B_{a}^{ij}$
and $\varphi \equiv \left( \varphi _{a},\varphi _{a}^{\left( 1\right)
},\varphi _{a}^{\left( 2\right) }\right) $ are not $\gamma _{I}$-invariant,
while the ghosts $\eta _{1a}^{i}$, $\eta _{1a}$, $\eta _{2a}$, together with
their spatial derivatives are trivial in the cohomology of $\gamma _{I}$ (as
they are $\gamma _{I}$-exact). The ghosts $\eta _{2a}^{i}$ and their
derivatives are $\gamma _{I}$-closed, but their antisymmetrized first-order
derivatives are $\gamma _{I}$-exact, as are also their subsequent
derivatives. Thus, the cohomology of $\gamma _{I}$ will be generated by $%
A_{i}^{a}$, $\partial _{i}B_{a}^{0i}$, $\pi \equiv \left( \pi _{ij}^{a},\pi
^{a},\pi ^{(1)a},\pi ^{(2)a}\right) $, $\mathcal{P}_{\Gamma }$, and their
spatial derivatives, as well as by the undifferentiated ghosts $\eta
_{2a}^{i}$ and their symmetrized first-order derivatives, $\partial ^{\left(
j\right. }\eta _{2a}^{\left. i\right) }$. Consequently, the general solution
of the equation $\gamma _{I}a=0$, can be written as 
\begin{equation}
a=a_{M}\left( \left[ A_{i}^{a}\right] ,\left[ \partial _{i}B_{a}^{0i}\right]
,\left[ \pi \right] ,\left[ \mathcal{P}_{\Gamma }\right] \right) e^{M}\left(
\eta _{2a}^{i},\partial ^{\left( j\right. }\eta _{2a}^{\left. i\right)
}\right) +\gamma _{I}b,  \label{27x}
\end{equation}
where $e^{M}\left( \eta _{2a}^{i},\partial ^{\left( j\right. }\eta
_{2a}^{\left. i\right) }\right) $ constitutes a basis in the
(finite-dimensional) space of the polynomials in the ghosts $\eta _{2a}^{i}$
and their symmetrized first-order derivatives, while the notation $a\left[
q\right] $ means that $a$ depends on $q$ and its spatial derivatives up to a
finite order. As $\mathrm{antigh}\left( \stackrel{\left( J\right) }{\omega }%
_{1}\right) =J$ and $\mathrm{gh}\left( \stackrel{\left( J\right) }{\omega }%
_{1}\right) =1$, we have that $\mathrm{pgh}\left( \stackrel{\left( J\right) 
}{\omega }_{1}\right) =J+1$. From (\ref{27x}) it results that the solution
to the equation $\gamma _{I}\stackrel{\left( J\right) }{\omega }_{1}=0$ is
(up to a trivial term) 
\begin{equation}
\stackrel{\left( J\right) }{\omega }_{1}=a_{J}\left( \left[ A_{i}^{a}\right]
,\left[ \partial _{i}B_{a}^{0i}\right] ,\left[ \pi \right] ,\left[ \mathcal{P%
}_{\Gamma }\right] \right) e^{J+1}\left( \eta _{2a}^{i},\partial ^{\left(
j\right. }\eta _{2a}^{\left. i\right) }\right) ,  \label{29x}
\end{equation}
where $\mathrm{antigh}\left( a_{J}\right) =J$. The equation (\ref{28x})
projected on antighost number $\left( J-1\right) $ takes the form 
\begin{equation}
\delta _{I}\stackrel{\left( J\right) }{\omega }_{1}+\gamma _{I}\stackrel{%
\left( J-1\right) }{\omega }_{1}=\partial _{i}\sigma ^{i}.  \label{30x}
\end{equation}
In order to ensure a solution to (\ref{30x}) (or, in other words, the
existence of $\stackrel{\left( J-1\right) }{\omega }_{1}$), it is necessary
that $a_{J}$ belongs to $H_{J}\left( \delta _{I}|\tilde{d}\right) $. With
the help of the Lagrangian results \cite{31} translated at the Hamiltonian
level, it follows that the cohomology of $\delta _{I}$ modulo $\tilde{d}$ is
vanishing in the case of the model under study for all antighost numbers
strictly greater that one, namely, 
\begin{equation}
H_{J}\left( \delta _{I}|\tilde{d}\right) =0,\;\mathrm{for}\;J>1.  \label{31x}
\end{equation}
In the meantime, the general representative of $H_{1}\left( \delta _{I}|%
\tilde{d}\right) $ can be written as 
\begin{equation}
\lambda =\lambda _{\;a}^{i}\mathcal{P}_{2i}^{a},  \label{32x}
\end{equation}
with $\lambda _{\;a}^{i}$ some constants, such that 
\begin{equation}
\delta _{I}\lambda =\partial ^{j}\left( \lambda _{\;a}^{i}\left( -\epsilon
_{0ijk}A^{ka}+g_{ij}\pi ^{a}\right) \right) .  \label{33x}
\end{equation}
On behalf of (\ref{31x}), it follows simply that the expansion of $\omega
_{1}$ stops after the first two terms, $\omega _{1}=\stackrel{\left(
0\right) }{\omega }_{1}+\stackrel{\left( 1\right) }{\omega }_{1}$, where $%
\stackrel{\left( 1\right) }{\omega }_{1}=a_{1}e^{2}\left( \eta
_{2a}^{i},\partial ^{\left( j\right. }\eta _{2a}^{\left. i\right) }\right) $%
, and $a_{1}$ pertains to $H_{1}\left( \delta _{I}|\tilde{d}\right) $.
Taking into account that $\mathrm{pgh}\left( e^{2}\left( \eta
_{2a}^{i},\partial ^{\left( j\right. }\eta _{2a}^{\left. i\right) }\right)
\right) =2$, there are only three possibilities, namely, $\eta _{2a}^{i}\eta
_{2b}^{j}$, $\eta _{2a}^{i}\partial ^{\left( j\right. }\eta _{2b}^{\left.
k\right) }$ and $\partial ^{\left( i\right. }\eta _{2a}^{\left. j\right)
}\partial ^{\left( k\right. }\eta _{2b}^{\left. l\right) }$, such that 
\begin{equation}
\stackrel{\left( 1\right) }{\omega }_{1}=a_{\;\;ij}^{ab}\eta _{2a}^{i}\eta
_{2b}^{j}+a_{\;\;ijk}^{ab}\eta _{2a}^{i}\partial ^{\left( j\right. }\eta
_{2b}^{\left. k\right) }+a_{\;\;ijkl}^{ab}\partial ^{\left( i\right. }\eta
_{2a}^{\left. j\right) }\partial ^{\left( k\right. }\eta _{2b}^{\left.
l\right) },  \label{33xa}
\end{equation}
with $a_{\;\;ij}^{ab}$, $a_{\;\;ijk}^{ab}$ and $a_{\;\;ijkl}^{ab}$ from $%
H_{1}\left( \delta _{I}|\tilde{d}\right) $, hence linear combinations of $%
\mathcal{P}_{2i}^{a}$ (see (\ref{32x})) 
\begin{equation}
a_{\;\;ij}^{ab}=a_{\;\;ijc}^{abk}\mathcal{P}_{2k}^{c},\;a_{\;\;ijk}^{ab}=a_{%
\;\;ijkc}^{abl}\mathcal{P}_{2l}^{c},\;a_{\;\;ijkl}^{ab}=a_{\;\;ijklc}^{abi^{%
\prime }}\mathcal{P}_{2i^{\prime }}^{c},  \label{34x}
\end{equation}
where $a_{\;\;ijc}^{abk}$, $a_{\;\;ijkc}^{abl}$ and $a_{\;\;ijklc}^{abi^{%
\prime }}$ are constants. Due to the covariance, these constants must
vanish, $a_{\;\;ijc}^{abk}=0$, $a_{\;\;ijkc}^{abl}=0$, $a_{\;\;ijklc}^{abi^{%
\prime }}=0$, so $\stackrel{\left( 1\right) }{\omega }_{1}=0$. So far, we
deduced that the only nonvanishing piece of $\omega _{1}$ is given by 
\begin{equation}
\stackrel{\left( 0\right) }{\omega }_{1}=a_{\;\;i}^{a}\left( \left[
A_{i}^{a}\right] ,\left[ \partial _{i}B_{a}^{0i}\right] ,\left[ \pi \right]
\right) \eta _{2a}^{i}.  \label{35x}
\end{equation}
Strictly speaking, $\stackrel{\left( 0\right) }{\omega }_{1}$ should have
contained also a term of the type $a_{\;\;ij}^{a}\partial ^{\left( i\right.
}\eta _{2a}^{\left. j\right) }$, with $a_{\;\;ij}^{a}=a_{\;\;ji}^{a}$. This
term can be rewritten under the form $a_{\;\;ij}^{a}\partial ^{\left(
i\right. }\eta _{2a}^{\left. j\right) }=\partial ^{i}\left(
2a_{\;\;ij}^{a}\eta _{2a}^{j}\right) -2\left( \partial
^{j}a_{\;\;ij}^{a}\right) \eta _{2a}^{i}$. As the non-integrated density $%
\stackrel{\left( 0\right) }{\omega }_{1}$ is defined up to a total
divergence, we can omit the term $\partial ^{i}\left( 2a_{\;\;ij}^{a}\eta
_{2a}^{j}\right) $, while the piece $-2\left( \partial
^{j}a_{\;\;ij}^{a}\right) \eta _{2a}^{i}$ is of the same type like that
appearing in (\ref{35x}) modulo the identification $a_{\;\;i}^{a}=-2\partial
^{j}a_{\;\;ij}^{a}$.

Let us investigate now the first-order deformation of the BRST-invariant
Hamiltonian. On account of (\ref{16x}) and (\ref{35x}), we find that 
\begin{eqnarray}
&&\left[ H_{0_{B}},\Omega _{1}\right] ^{*}=\int d^{3}x\left(
-a_{\;\;i}^{a}\left( \eta _{1a}^{i}+\partial ^{i}\eta _{2a}\right) +\right. 
\nonumber \\
&&\int d^{3}y\eta _{2b}^{i}\left( y\right) \left( \left[ \varphi _{a}\left(
x\right) ,a_{\;\;i}^{b}\left( y\right) \right] ^{*}\pi ^{(2)a}\left(
x\right) -\left[ \varphi _{a}^{\left( 2\right) }\left( x\right)
,a_{\;\;i}^{b}\left( y\right) \right] ^{*}\partial _{i}\partial ^{i}\pi
^{a}\left( x\right) -\right.  \nonumber \\
&&\left. \left. \left[ B_{a}^{0j}\left( x\right) ,a_{\;\;i}^{b}\left(
y\right) \right] ^{*}\partial _{j}\partial ^{k}B_{0k}^{a}\left( x\right)
\right) \right) .  \label{36x}
\end{eqnarray}
In order to ensure the compensation of the term $a_{\;\;i}^{a}\eta _{1a}^{i}$
through a similar term in $\left[ H_{1},\Omega _{0}\right] ^{*}$, we take $%
H_{1}$ of the type 
\begin{equation}
H_{1}=\int d^{3}x\left( -\frac{1}{2}\epsilon ^{0ijk}a_{\;\;i}^{a}B_{ajk}+%
\bar{h}_{1}\right) ,  \label{37x}
\end{equation}
where $\bar{h}_{1}$ does not involve $B_{ajk}$. By means of (\ref{37x}), it
results that 
\begin{eqnarray}
& &\left[ H_{1},\Omega _{0}\right] ^{*}=\int d^{3}x\left( a_{\;\;i}^{a}\eta
_{1a}^{i}-\right.  \nonumber \\
& &\int d^{3}y\left( \frac{1}{4}\epsilon ^{0ijk}\epsilon
_{0lmn}B_{ajk}\left( x\right) \left[ a_{\;\;i}^{a}\left( x\right)
,F^{bmn}\left( y\right) \right] ^{*}\eta _{2b}^{l}\left( y\right) -\right. 
\nonumber \\
& &\left. \left. \left[ \bar{h}_{1}\left( x\right) ,\omega _{0}\left(
y\right) \right] ^{*}\right) \right) ,  \label{38x}
\end{eqnarray}
such that the first equation in (\ref{26x}) becomes 
\begin{eqnarray}
&&-\int d^{3}xa_{\;\;i}^{a}\partial ^{i}\eta _{2a}+\int d^{3}xd^{3}y\left(
\eta _{2b}^{l}\left( y\right) \left( \left[ \varphi _{a}\left( x\right)
,a_{\;\;l}^{b}\left( y\right) \right] ^{*}\pi ^{(2)a}\left( x\right)
-\right. \right.  \nonumber \\
&&\left[ \varphi _{a}^{\left( 2\right) }\left( x\right) ,a_{\;\;l}^{b}\left(
y\right) \right] ^{*}\partial _{i}\partial ^{i}\pi ^{a}\left( x\right)
-\left[ B_{a}^{0j}\left( x\right) ,a_{\;\;l}^{b}\left( y\right) \right]
^{*}\partial _{j}\partial ^{k}B_{0k}^{a}\left( x\right) -  \nonumber \\
&&\left. \left. \frac{1}{4}\epsilon ^{0ijk}\epsilon _{0lmn}B_{ajk}\left(
x\right) \left[ a_{\;\;i}^{a}\left( x\right) ,F^{bmn}\left( y\right) \right]
^{*}\right) +\left[ \bar{h}_{1}\left( x\right) ,\omega _{0}\left( y\right)
\right] ^{*}\right) =0.  \label{39x}
\end{eqnarray}
The last equation is verified with the choices 
\begin{equation}
a_{\;\;i}^{a}=-f_{\;\;bc}^{a}\left( \frac{1}{2}\epsilon
_{0ijk}A^{bj}A^{ck}+A_{i}^{c}\pi ^{b}\right) ,  \label{40x}
\end{equation}
\begin{eqnarray}
&&\bar{h}_{1}=f_{\;\;bc}^{a}\left( -\left( \partial ^{i}B_{0i}^{b}\right)
\left( B_{a}^{0j}A_{j}^{c}+\varphi _{a}\pi ^{c}+\varphi _{a}^{\left(
1\right) }\pi ^{c}+\varphi _{a}^{\left( 2\right) }\pi ^{\left( 2\right)
c}+\right. \right.  \nonumber \\
&&\left. \left. \eta _{2a}^{j}\mathcal{P}_{2j}^{c}+\eta _{2a}\mathcal{P}%
_{2}^{c}\right) +A_{i}^{c}\left( \pi ^{b}\partial ^{i}\varphi _{a}^{\left(
2\right) }-\eta _{2a}\mathcal{P}_{2}^{bi}+\eta _{2a}^{i}\mathcal{P}%
_{2}^{b}\right) \right) ,  \label{41x}
\end{eqnarray}
where $f_{\;\;bc}^{a}$ are some constants, antisymmetric in the lower
indices, $f_{\;\;bc}^{a}=-f_{\;\;cb}^{a}$. Inserting (\ref{40x}-\ref{41x})
in (\ref{35x}) and (\ref{37x}), we determine the complete expressions of the
first-order deformations of both BRST charge and BRST-invariant Hamiltonian.

Next, we analyse the second-order deformations. With $\Omega _{1}$ at hand,
we find that $\left[ \Omega _{1},\Omega _{1}\right] ^{*}=0$, so the second
equation in (\ref{25x}) is obeyed for $\Omega _{2}=0$. The higher-order
equations that describe the deformation of the BRST charge will be satisfied
if we set $\Omega _{3}=\Omega _{4}=\cdots =0$. On the one hand, as $\Omega
_{2}=0$, the second equation in (\ref{26x}) reduces to 
\begin{equation}
\left[ H_{2},\Omega _{0}\right] ^{*}+\left[ H_{1},\Omega _{1}\right] ^{*}=0.
\label{42x}
\end{equation}
On the other hand, $\left[ H_{1},\Omega _{1}\right] ^{*}$ is given by 
\begin{eqnarray}
& &\left[ H_{1},\Omega _{1}\right] ^{*}=s_{I}\left( \int d^{3}x\left(
f_{\;\;bc}^{a}f_{\;\;ea}^{d}A_{i}^{c}\pi ^{b}A^{ei}\varphi _{d}^{\left(
2\right) }+\right. \right.  \nonumber \\
& &f_{\;\;bc}^{a}f_{\;\;eg}^{d}g^{eb}\left( \left( \varphi _{a}+\varphi
_{a}^{\left( 1\right) }\right) \pi ^{c}\left( \frac{1}{2}\left( \varphi
_{d}+\varphi _{d}^{\left( 1\right) }\right) \pi
^{g}+B_{d}^{0j}A_{j}^{g}+\right. \right.  \nonumber \\
& &\left. \varphi _{d}^{\left( 2\right) }\pi ^{\left( 2\right) g}+ \eta
_{2d}^{j}\mathcal{P}_{2j}^{g}+\eta _{2d}\mathcal{P}_{2}^{g}\right)
+B_{a}^{0i}A_{i}^{c}\left( \frac{1}{2}B_{d}^{0j}A_{j}^{g}+\varphi
_{d}^{\left( 2\right) }\pi ^{\left( 2\right) g}+\right.  \nonumber \\
& &\left. \left. \left. \left. \eta _{2d}\mathcal{P}_{2}^{g}\right) + \eta
_{2a}^{i}\mathcal{P}_{2i}^{c}\left( B_{d}^{0j}A_{j}^{g}+\varphi _{d}^{\left(
2\right) }\pi ^{\left( 2\right) g}+\frac{1}{2}\eta _{2d}^{j}\mathcal{P}%
_{2j}^{g}+\eta _{2d}\mathcal{P}_{2}^{g}\right) \right) \right) \right) - 
\nonumber \\
& &f_{\;\;\left[ de\right. }^{c} f_{\;\;\left. b\right] c}^{a}\int
d^{3}x\left( \eta _{2a}^{i}\partial ^{l}B_{0l}^{d}\left( A_{i}^{e}\pi ^{b}+%
\frac{1}{2}\epsilon _{0ijk}A^{bj}A^{ek}\right) +\right.  \nonumber \\
& &\left. \frac{1}{3}\epsilon ^{0ijk}\eta
_{2a}A_{i}^{d}A_{j}^{e}A_{k}^{b}\right) ,  \label{43x}
\end{eqnarray}
where $\left[ deb\right] $ means antisymmetry with respect to the indices
between brackets. From (\ref{43x}), we remark that $\left[ H_{1},\Omega
_{1}\right] ^{*}$ is indeed $s_{I}$-exact (as required by (\ref{42x})) if
and only if the antisymmetric constants $f_{\;\;bc}^{a}$ fulfill the Jacobi
identity 
\begin{equation}
f_{\;\;\left[ de\right. }^{c}f_{\;\;\left. b\right] c}^{a}=0,  \label{44x}
\end{equation}
therefore if and only if they stand for the structure constants of a Lie
algebra. Under these circumstances, we arrive at 
\begin{eqnarray}
&&h_{2}=f_{\;\;bc}^{a}f_{\;\;ea}^{d}A_{i}^{c}\pi ^{b}A^{ei}\varphi
_{d}^{\left( 2\right) }+f_{\;\;bc}^{a}f_{\;\;eg}^{d}g^{eb}\left( \left(
\varphi _{a}+\varphi _{a}^{\left( 1\right) }\right) \pi ^{c}\left( \frac{1}{2%
}\left( \varphi _{d}+\varphi _{d}^{\left( 1\right) }\right) \pi ^{g}+\right.
\right.  \nonumber \\
&&\left. B_{d}^{0j}A_{j}^{g}+\varphi _{d}^{\left( 2\right) }\pi ^{\left(
2\right) g}+\eta _{2d}^{j}\mathcal{P}_{2j}^{g}+\eta _{2d}\mathcal{P}%
_{2}^{g}\right) +B_{a}^{0i}A_{i}^{c}\left( \frac{1}{2}B_{d}^{0j}A_{j}^{g}+%
\varphi _{d}^{\left( 2\right) }\pi ^{\left( 2\right) g}+\right.  \nonumber \\
&&\left. \left. \eta _{2d}\mathcal{P}_{2}^{g}\right) +\eta _{2a}^{i}\mathcal{%
P}_{2i}^{c}\left( B_{d}^{0j}A_{j}^{g}+\varphi _{d}^{\left( 2\right) }\pi
^{\left( 2\right) g}+\frac{1}{2}\eta _{2d}^{j}\mathcal{P}_{2j}^{g}+\eta _{2d}%
\mathcal{P}_{2}^{g}\right) \right) .  \label{45x}
\end{eqnarray}
The equation that governs the third-order deformation of the BRST-invariant
Hamiltonian 
\begin{equation}
\left[ H_{3},\Omega _{0}\right] ^{*}+\left[ H_{2},\Omega _{1}\right]
^{*}+\left[ H_{1},\Omega _{2}\right] ^{*}+\left[ H_{0},\Omega _{3}\right]
^{*}=0,  \label{46x}
\end{equation}
simply becomes $\left[ H_{3},\Omega _{0}\right] ^{*}+\left[ H_{2},\Omega
_{1}\right] ^{*}=0$, because $\Omega _{2}=\Omega _{3}=0$. Using (\ref{45x}),
we find that $\left[ H_{2},\Omega _{1}\right] ^{*}=0$ due to the Jacobi
identity. Accordingly, we can take $H_{3}=0$, and similarly for the
higher-order deformation equations, which are satisfied for $%
H_{4}=H_{5}=\cdots =0$. In this way, the Hamiltonian deformation procedure
in the context of the irreducible BRST formulation of the model under study
is completed.

\section{Identification of the deformed model}

In the sequel we analyse the deformed theory constructed in the above.
Collecting all the results deduced so far, we can write down the complete
BRST charge and BRST-invariant Hamiltonian of the deformed model, that are
consistent to all orders in the deformation parameter, under the form 
\begin{eqnarray}
&&\Omega =\int d^{3}x\left( \epsilon _{0ijk}\pi ^{jka}\eta _{1a}^{i}+\left( 
\frac{1}{2}\epsilon _{0ijk}H^{ajk}-\left( D_{i}\right) _{\;\;b}^{a}\pi
^{b}\right) \eta _{2a}^{i}+\right.  \nonumber \\
&&\left. \left( \pi ^{a}-\pi ^{(1)a}\right) \eta _{1a}-\pi ^{(2)a}\eta
_{2a}\right) ,  \label{47x}
\end{eqnarray}
respectively, 
\begin{eqnarray}
& &H_{B}=\int d^{3}x\left( \frac{1}{2}B_{a}^{ij}\left( H_{ij}^{a}+
\varepsilon _{0ijk}\left( D^{k}\right) _{\;\;b}^{a}\pi ^{b}\right) -\frac{1}{%
2}A_{i}^{a}A_{a}^{i}+\varphi _{a}\pi ^{(2)a}+\right.  \nonumber \\
& &\frac{1}{2}\left( \left( D_{i}\right)
_{a}^{\;\;b}B_{b}^{0i}-gf_{\;\;ab}^{c}\left( \varphi _{c}\pi ^{b}+\varphi
_{c}^{(1)}\pi ^{b}+\varphi _{c}^{(2)}\pi ^{(2)b}\right) \right) ^{2}- 
\nonumber \\
& &\varphi _{a}^{(2)}\left( \left( D_{i}\right) _{\;\;b}^{a} \left(
D^{i}\right) _{\;\;c}^{b}\right) \pi ^{c}+\eta _{1a}^{i}\mathcal{P}%
_{2i}^{a}-\eta _{1a}\mathcal{P}_{2}^{a}-\eta _{2a}\left( D^{i}\right)
_{\;\;b}^{a}\mathcal{P}_{2i}^{b}+  \nonumber \\
& &\eta _{2a}^{i}\left( D_{i}\right) _{\;\;b}^{a}\mathcal{P}_{2}^{b}-
gf_{\;\;ab}^{c}\left( \eta _{2c}^{i}\mathcal{P}_{2i}^{b}+\eta _{2c}\mathcal{P%
}_{2}^{b}\right) \times  \nonumber \\
& &\left( \left( D^{j}\right) _{\;\;d}^{a}B_{0j}^{d}- gf_{\;\;de}^{a}\left(
\pi ^{d}\varphi ^{e}+\pi ^{d}\varphi ^{(1)e}+\pi ^{(2)d}\varphi
^{(2)e}\right) \right) -  \nonumber \\
& &\left. \frac{1}{2}g^{2}f_{\;\;ab}^{c}f_{\;\;de}^{a}\left( \eta _{2c}^{i}%
\mathcal{P}_{2i}^{b}+\eta _{2c}\mathcal{P}_{2}^{b}\right) \left( \eta
_{2j}^{d}\mathcal{P}_{2}^{je}+\eta _{2}^{d}\mathcal{P}_{2}^{e}\right)
\right) ,  \label{48x}
\end{eqnarray}
where we employed the notations 
\begin{equation}
H_{ij}^{a}=F_{ij}^{a}-gf_{\;\;bc}^{a}A_{i}^{b}A_{j}^{c},  \label{49x}
\end{equation}
\begin{eqnarray}
&&\left( \left( D_{i}\right) _{a}^{\;\;b}B_{b}^{0i}-gf_{\;\;ab}^{c}\left(
\varphi _{c}\pi ^{b}+\varphi _{c}^{(1)}\pi ^{b}+\varphi _{c}^{(2)}\pi
^{(2)b}\right) \right) ^{2}\equiv  \nonumber \\
&&\left( \left( D_{i}\right) _{a}^{\;\;b}B_{b}^{0i}-gf_{\;\;ab}^{c}\left(
\varphi _{c}\pi ^{b}+\varphi _{c}^{(1)}\pi ^{b}+\varphi _{c}^{(2)}\pi
^{(2)b}\right) \right) \times  \nonumber \\
&&\left( \left( D^{j}\right) _{\;\;d}^{a}B_{0j}^{d}-gf_{\;\;de}^{a}\left(
\pi ^{d}\varphi ^{e}+\pi ^{d}\varphi ^{(1)e}+\pi ^{(2)d}\varphi
^{(2)e}\right) \right) ,  \label{50x}
\end{eqnarray}
\begin{equation}
\left( D_{i}\right) _{\;\;b}^{a}=\delta _{\;\;b}^{a}\partial
_{i}+gf_{\;\;bc}^{a}A_{i}^{c},\;\left( D_{i}\right) _{b}^{\;\;a}=\delta
_{b}^{\;\;a}\partial _{i}-gf_{\;\;bc}^{a}A_{i}^{c}.  \label{51x}
\end{equation}
From the terms of antighost number zero in (\ref{47x}) we read that only the
first-class constraints $\gamma _{2i}^{a}\equiv \frac{1}{2}\epsilon
_{0ijk}F^{ajk}-\partial _{i}\pi ^{a}\approx 0$ are deformed like 
\begin{equation}
\bar{\gamma}_{2i}^{a}\equiv \frac{1}{2}\epsilon _{0ijk}H^{jka}-\left(
D_{i}\right) _{\;\;b}^{a}\pi ^{b}\approx 0,  \label{52x}
\end{equation}
while the others are kept unchanged. In the meantime, the resulting BRST
charge contains no pieces quadratic in the ghost number one ghosts and
linear in the antighosts, hence the gauge algebra (in the Dirac bracket) of
the deformed first-class constraints remains abelian, being not affected by
the deformation method. Analysing the structure of the pieces in (\ref{48x})
that involve neither ghosts, nor antighosts, we discover that the
first-class Hamiltonian of the deformed theory reads as 
\begin{eqnarray}
&&H=\int d^{3}x\left( \frac{1}{2}B_{a}^{ij}\left( H_{ij}^{a}+\varepsilon
_{0ijk}\left( D^{k}\right) _{\;\;b}^{a}\pi ^{b}\right) -\frac{1}{2}%
A_{i}^{a}A_{a}^{i}+\varphi _{a}\pi ^{(2)a}+\right.  \nonumber \\
&&\frac{1}{2}\left( \left( D_{i}\right)
_{a}^{\;\;b}B_{b}^{0i}-gf_{\;\;ab}^{c}\left( \varphi _{c}\pi ^{b}+\varphi
_{c}^{(1)}\pi ^{b}+\varphi _{c}^{(2)}\pi ^{(2)b}\right) \right) ^{2}- 
\nonumber \\
&&\left. \varphi _{a}^{(2)}\left( \left( D_{i}\right) _{\;\;b}^{a}\left(
D^{i}\right) _{\;\;c}^{b}\right) \pi ^{c}\right) ,  \label{53x}
\end{eqnarray}
while from the components linear in the antighost number one antighosts we
find that the Dirac brackets among the new first-class Hamiltonian and the
new first-class constraints are modified as 
\begin{equation}
\left[ H,\gamma _{1i}^{a}\right] ^{*}=\bar{\gamma}_{2i}^{a},\;\left[
H,\gamma _{1}^{a}\right] ^{*}=-\gamma _{2}^{a},  \label{54x}
\end{equation}
\begin{eqnarray}
&&\left[ H,\bar{\gamma}_{2i}^{a}\right] ^{*}=\left( D_{i}\right)
_{\;\;b}^{a}\gamma _{2}^{b}-  \nonumber \\
&&gf_{bc}^{a}\left( \left( D^{j}\right)
_{\;\;d}^{b}B_{0j}^{d}-gf_{de}^{b}\left( \pi ^{d}\varphi ^{e}+\pi
^{d}\varphi ^{(1)e}+\pi ^{(2)d}\varphi ^{(2)e}\right) \right) \bar{\gamma}%
_{2i}^{c},  \label{55x}
\end{eqnarray}
\begin{eqnarray}
&&\left[ H,\gamma _{2}^{a}\right] ^{*}=-\left( D^{i}\right) _{\;\;b}^{a}\bar{%
\gamma}_{2i}^{b}-  \nonumber \\
&&gf_{bc}^{a}\left( \left( D^{j}\right)
_{\;\;d}^{b}B_{0j}^{d}-gf_{de}^{b}\left( \pi ^{d}\varphi ^{e}+\pi
^{d}\varphi ^{(1)e}+\pi ^{(2)d}\varphi ^{(2)e}\right) \right) \gamma
_{2}^{c}.  \label{56x}
\end{eqnarray}
The first-class constraints and first-class Hamiltonian generated until now
along the deformation scheme reveal precisely the consistent irreducible
Hamiltonian interactions that can be introduced among a set of two-form
gauge fields, which actually produce the irreducible version of non-abelian
Freedman-Townsend model in four-dimensions. As the first-class constraints
generate gauge transformations, we can state that the added interactions
deform the gauge transformations, but not the algebra of gauge
transformations (due to the abelianity of the deformed first-class
constraints). In order to make our result more transparent, we observe that
if in (\ref{53x}--\ref{56x}) we set all the supplementary canonical pairs $%
\left( \varphi _{a},\pi ^{a}\right) $, $\left( \varphi _{a}^{(1)},\pi
^{(1)a}\right) $, and $\left( \varphi _{a}^{(2)},\pi ^{(2)a}\right) $ equal
to zero, we reach that the Lagrangian form of the interaction term generated
by our procedure is precisely $\frac{1}{2}f_{\;bc}^{a}B_{a}^{\mu \nu }A_{\mu
}^{b}A_{\nu }^{c}$, which is nothing but the standard Freedman-Townsend
vertex, while the remaining gauge algebra relations reduce to those of the
usual Freedman-Townsend model. This completes our analysis.

\section{Conclusion}

To conclude with, in this paper we have investigated in an irreducible
manner the consistent Hamiltonian interactions among a set of two-forms in
four dimensions. Our method is based on the deformation of the BRST charge
and BRST-invariant Hamiltonian associated with an irreducible formulation of
the free model. The main equations that control the deformed quantities are
solved by using some cohomological techniques. As a result, we obtain the
irreducible version of the non-abelian Freedman-Townsend model. The added
interactions deform the gauge transformations (but not their algebra), as
well as the Dirac brackets between the first-class Hamiltonian and the
first-class constraint set. The interaction term revealed by our method is
nothing but the standard Freedman-Townsend vertex.

\end{document}